\documentclass[aps,prl,twocolumn,groupedaddress,showpacs]{revtex4}

\usepackage{bm}
\usepackage{graphicx}

\begin{document}

\title{Direct Evidence for the Dirac-Cone Topological Surface States in Ternary Chalcogenide TlBiSe$_2$}
\author{Takafumi Sato,$^1$ Kouji Segawa,$^2$ Hua Guo,$^1$ Katsuaki Sugawara,$^3$ Seigo Souma,$^3$ Takashi Takahashi,$^{1,3}$ and Yoichi Ando$^2$}
\affiliation{$^1$Department of Physics, Tohoku University, Sendai 980-8578, Japan}
\affiliation{$^2$Institute of Scientific and Industrial Research, Osaka University, Ibaraki, Osaka 567-0047, Japan}
\affiliation{$^3$WPI Research Center, Advanced Institute for Materials Research, 
Tohoku University, Sendai 980-8577, Japan}
\date{\today}

\begin{abstract}	
We have performed angle-resolved photoemission spectroscopy on TlBiSe$_2$, which is a member of the ternary chalcogenides theoretically proposed as candidates for a new class of three-dimensional topological insulators (TIs).  We found a direct evidence for a non-trivial surface metallic state showing a \,`X\,'-shaped energy dispersion within the bulk band gap.  The present result unambiguously establishes that TlBiSe$_2$ is a strong TI with a single Dirac cone at the Brillouin-zone center.  The observed bulk band gap of 0.35 eV is the largest among known TIs, making TlBiSe$_2$ the most promising material for studying room-temperature topological phenomena.
\end{abstract}

\maketitle

 Topological insulators (TIs) are recently attracting significant attentions, since they materialize a new state of matter where the bulk excitation gap generated by the spin-orbit coupling (SOC) leads to the appearance of unusual metallic states at the edge or surface due to a topological principle.  A two-dimensional (2D) form of TIs was first realized in HgTe quantum wells \cite{HgTe_theory, HgTe_exp}, for which the observation of a quantized conductivity associated with the one-dimensional (1D) helical edge states gave evidence.  In analogy with the 1D edge state in the 2D TIs, a 2D surface state (SS) appears in three-dimensional (3D) TIs.  Such a SS emerges within the bulk band gap and exhibits a gapless band dispersion protected by the time reversal symmetry (TRS) \cite{KanePRL}.   This novel SS has been the subject of intensive investigations because of not only its fundamental novelty but also its high potential for low-power spintronics devices \cite{KaneScience} and fault-tolerant quantum computations \cite{FuKanePRL}.  In fact, the topological 2D SS has been observed in group-V alloy Bi$_{1-x}$Sb$_x$ (which becomes a narrow-gap semiconductor for 0.07 $<$ $x$ $<$ 0.22) \cite{FuKanePRB, JCYTeoPRB, HasanBiSbNature, HasanBiSbScience, TaskinPRB, MatsudaPRB} as well as in tetradymite semiconductors Bi$_2$Se$_3$ and Bi$_2$Te$_3$ \cite{Hasan23, HasanCa, Kim23, Shen23, calc23}.  
 
Angle-resolved photoemission spectroscopy (ARPES) has played a central role in the identification of TIs by utilizing its unique capability to determine the momentum ($k$) resolved electronic states.  ARPES studies of Bi$_{1-x}$Sb$_x$ have revealed the complex nature of topological SS where five or three (depending on $x$) energy bands cross the Fermi level ($E_{\rm F}$) along the $\bar{\Gamma}$$\bar{\rm{M}}$ line \cite{HasanBiSbNature,HasanBiSbScience,MatsudaPRB}, while ARPES studies of Bi$_2$Se$_3$ \cite{Hasan23,HasanCa,Kim23} and Bi$_2$Te$_3$ \cite{Shen23} uncovered simpler topological SS consisting of a single Dirac cone at the $\bar{\Gamma}$ point.  From these ARPES studies, all of these materials were categorized as the strong TIs characterized by the topological invariant ($Z$$_2$ number) of $\nu$$_0$ = 1 \cite{FuKanePRB,Hasan_review}.
 
Recently, it has been proposed by the band structure calculations from two independent groups \cite{Tl112SCZhang, Tl112Banzil} that thallium-based III-V-IV$_2$ ternary chalcogenides TlM\,'X$_2$ (M\,'= Bi and Sb; X = S, Se, and Te) would be candidates for a new class of 3D TIs with a single Dirac cone at the $\bar{\Gamma}$ point \cite{Tl112Banzil, Tl112SCZhang}.  However, a clear experimental demonstration of the TI nature of this system has not yet been made.  Finding a new class of TIs with a large band gap is particularly important, since only a few TI materials have been found so far and each has its own chemical reason to make it difficult to obtain a bulk-insulating sample, which is a prerequisite to studying peculiar surface transport properties.

In this Letter, we report an ARPES study of TlBiSe$_2$.  Our observation of a single Dirac cone SS within the large bulk band gap at the zone center provides a direct experimental evidence that this materials is a promising new 3D TI, because the observed band gap of 0.35 eV is the largest among known TIs.  We discuss the present ARPES results in relation to the band calculations as well as to previous ARPES studies of other TIs.

Single crystals of TlBiSe$_2$ were grown by the Bridgman method which was modified from that previously reported \cite{TlBiSe2_Growth}.  The purity of starting materials is 99.9999$\%$ for Bi and Se, and 99.999$\%$ for Tl, respectively.  We have determined the Tl, Bi, and Se contents as 0.99, 1.01, and 2.00, respectively, from the ICP-AES (inductively coupled plasma - atomic emission spectroscopy) analysis.  TlBiSe$_2$ has a rhombohedral crystal structure with the space group D$_{3d}^5$ ($R\bar{3}m$) which possesses the real-space inversion symmetry, and it can be viewed as distorted NaCl structure with four atoms in the primitive unit cell as shown in Fig. 1(a).  The stacking sequence of each layer is -Tl-Se-Bi-Se- along the [111] direction, and the binding between the layers is rather strong in contrast to the van-der-Waals type coupling of tetradymite semiconductors.  The bulk BZ and its projected surface BZ on the (111) plane are shown in Fig. 1(b).  

ARPES measurements were performed using VG-Scienta SES2002 and MBS-A1 spectrometers with high-flux He and Xe discharge lamps at Tohoku University.  The He I$\alpha$ ($h\nu$ = 21.218 eV) line and one of the Xe I ($h\nu$ = 8.437 eV) line \cite{SoumaRSI} were used to excite photoelectrons.  Samples were cleaved {\it in-situ} along the (111) crystal plane in an ultrahigh vacuum of 5$\times$10$^{-11}$ Torr.  The energy and angular resolutions were set at 2-15 meV and 0.2$^{\circ}$, respectively.  A part of the ARPES data has been obtained at the BL28 beamline in Photon Factory (PF), KEK, Tsukuba.  Transport characterization of the crystal was done using a standard six-probe method.

Figures 1(c) and (d) show valence-band (VB) ARPES spectra of TlBiSe$_2$ measured along $\bar{\Gamma}$$\bar{\rm{M}}$ and $\bar{\Gamma}$$\bar{\rm{K}}$ in the surface BZ.  Corresponding second-derivative intensity plots are also shown in Figs. 1(e) and (f).  As seen in Figs. 1(e) and (f), several dispersive bands are observed at higher than $\sim$0.7 eV.  These bands are attributed to the hybridized states of Tl/Bi 6$p$ and Se 4$p$ orbitals \cite{Tl112SCZhang, Tl112Banzil, HoangPRB}.  Apart from such a complicated band dispersion, a holelike band with the top at around 0.5-1.0 eV is clearly visible at the $\bar{\Gamma}$ point as shown by red arrows.  These states correspond to the top of the VB of predominantly Se 4$p$ character at the ${\Gamma}$ point of the bulk BZ \cite{HoangPRB}.  According to the band calculation, this bulk VB also shows a local maximum at the F point of the bulk BZ \cite{Tl112SCZhang, Tl112Banzil, HoangPRB}, and may correspond to the experimentally observed holelike band around the $\bar{\rm{M}}$ point, as indicated by yellow arrows.   In addition to these VB features, we find a weak intensity near $E_{\rm F}$ at the $\bar{\Gamma}$ point.  
\begin{figure}[hb]
\includegraphics[width=3.0in]{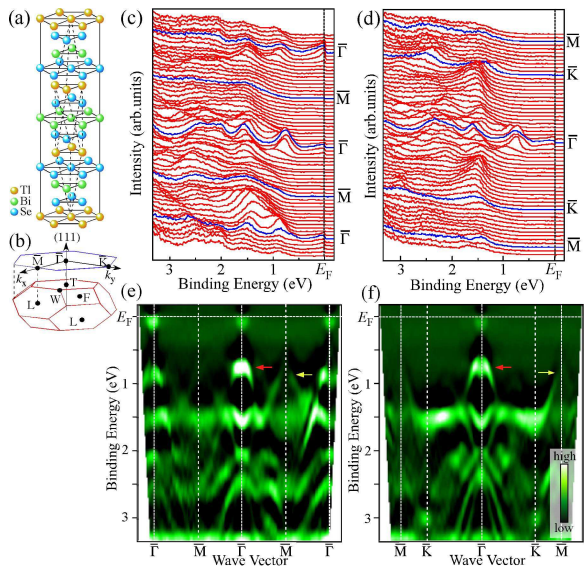}
\caption{(Color) (a) Crystal structure of TlBiSe$_2$. (b) Bulk BZ (red) and corresponding (111) surface BZ (blue).  (c),(d) ARPES spectra at $T$ = 30 K along $\bar{\Gamma}$$\bar{\rm{M}}$ and $\bar{\Gamma}$$\bar{\rm{K}}$ lines, respectively, measured with He I$\alpha$ line.  (e), (f) Second-derivative ARPES intensity of (c) and (d), respectively.
}
\end{figure}
This band is assigned to the bottom of the conduction band (CB) of predominantly Tl/Bi 6$p$ character, which lies above $E_{\rm F}$ in the calculations \cite{Tl112SCZhang, Tl112Banzil, HoangPRB}.  In the present ARPES experiment, this band is shifted downward to cross $E_{\rm F}$, possibly owing to the electron-doped character of naturally grown crystals as in the case of Bi$_2$Se$_3$ and Bi$_2$Te$_3$ \cite{Hasan23,  HasanCa, Shen23, HorPRB, CheckelskyPRL}.  In fact, the electrical resistivity and Hall-coefficient measurements of the same crystal show the metallic behavior and the electron density (carrier number) of about 5$\times$10$^{19}$ cm$^{-3}$, consistent with the observed metallic Fermi surface (FS) in the ARPES measurement.  It is inferred that the slight excess of Bi in the crystal is responsible for an electron doping to make the present sample metallic.

To estimate the electron carrier number from the ARPES data  \cite{SoumaCaB6}, we have determined the Fermi wave vector ($k_{\rm F}$) of the bulk CB as 0.11$\pm$0.02\AA$^{-1}$ by the momentum distribution curves near $E_{\rm F}$, and estimated the FS volume as 0.25$\pm$0.1$\%$ of the bulk BZ volume by assuming a spherical FS.  The estimated carrier number of 5$\pm$2$\times$10$^{19}$ cm$^{-3}$ is in good agreement with the bulk carrier number of 5$\times$10$^{19}$ cm$^{-3}$, indicating that observed ARPES spectra certainly reflect bulk properties.  It is also inferred that the surface band bending effect would be small because the carrier numbers estimated from the ARPES and the Hall-coefficient measurements agree fairly well.  As we discuss later, the electron doping turns out to be useful for elucidating the whole picture of the SS dispersion.

Figure 2 displays the ARPES intensity at $E_{\rm F}$ plotted as a function of 2D wave vector.  A FS centered at the $\bar{\Gamma}$ point is clearly visible in both the first and second BZs consistent with Fig. 1, while we do not find other FSs as confirmed by the FS mapping over the entire first BZ.  This is consistent with the bulk band calculations where the CB minimum is always located at the $\Gamma$ point in the bulk BZ \cite{HoangPRB, Tl112Banzil, Tl112SCZhang}.

To elucidate the near-$E_{\rm F}$ electronic states around the $\bar{\Gamma}$ point in detail, we have performed ARPES measurements with 
\begin{figure}[h]
\includegraphics[width=2.2in]{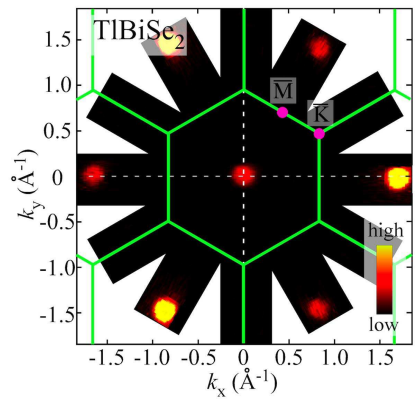}
\caption{(Color) ARPES intensity at $E_{\rm F}$ as a function of 2D wave vector.  The intensity is integrated over $\pm$20 meV with respect to $E_{\rm F}$, and symmetrized with the three-fold symmetry.}
\end{figure}
higher energy resolution and finer $k$ interval. As seen in the He spectrum in Fig. 3(a), a broad intense spectral feature is observed around 0.6-0.8 eV, above which the intensity is rapidly reduced.  This sudden weight suppression is due to the bulk band-gap opening.  The most important aspect of Fig. 3(a) is the presence of a X-shaped band dispersion reminiscent of a Dirac cone, which is more clearly seen in the Xe spectrum in Fig. 3(b).  We find in Fig. 3 that the bulk-band feature is remarkably different between the He and Xe measurements.  This is reasonably expected for the bulk bands since the momentum perpendicular to the crystal surface ($k_z$) is different for different photon energies.  On the other hand, the energy position of the X-shaped band is identical between the two measurements, giving compelling evidence for its surface origin.   As seen in Fig. 3(b), each branch emerges from the left- and right-side edge of the VB continuum at $\sim$0.6 eV [see also Fig. 3(a)], crosses each other at 0.43 eV at the $\bar{\Gamma}$ point, and finally merges into the CB at 0-0.2 eV.

To estimate the bulk band-gap size, we have performed ARPES measurements with the synchrotron radiation, and determined the band structure by varying $k_{\rm z}$.  From the normal-emission measurement ($\Gamma$T cut in the bulk BZ in Fig. 1(b)), the VB maximum and the CB minimum were found to be located around the $\Gamma$ point of the bulk BZ.  Also, the photon energy of 38 eV was found to be best suited for seeing the VB maximum.  As shown by the corresponding ARPES intensity 
\begin{figure}[h]
\includegraphics[width=2.8in]{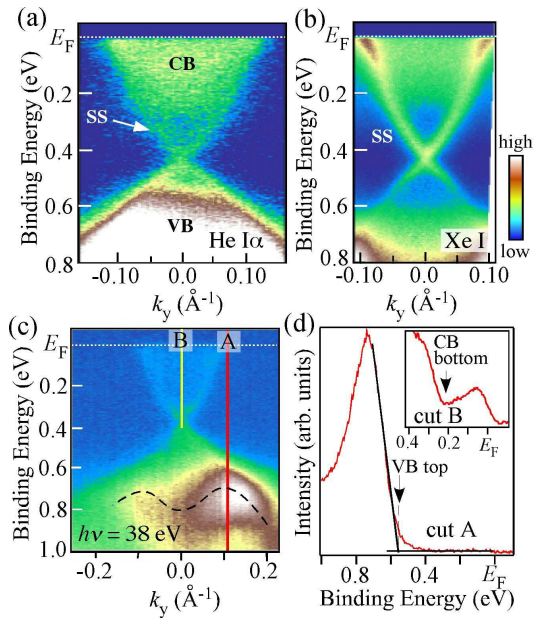}
\caption{(Color) Near-$E_{\rm F}$ ARPES intensity around the $\bar{\Gamma}$ point plotted as a function of $k_y$ and binding energy measured with (a) He I$\alpha$, (b) Xe I, and (c) $h\nu$ = 38 eV.  Black dashed line is a guide to the eye for tracing the VB dispersion.  (d) ARPES spectrum at the momentum indicated by a red line in (c) (cut A).  Inset shows the spectrum at cut B [yellow line in (c)].}
\end{figure}
plot at $h\nu$ = 38 eV in Fig. 3(c), the VB shows a characteristic M-shaped dispersion with the top of dispersion slightly away from the zone center, while the CB minimum is located at the $\Gamma$ point, suggesting an indirect band gap.  By determining the leading (trailing) edge of the VB (CB) as shown by an arrow in Fig. 3(d), we have determined the band gap size to be 0.35 eV.  This value is in accordance with that from the optical absorption experiment of TlBiSe$_2$ thin film ($\sim$0.45 eV) \cite{Optical}, and is larger than that of tetradymite semiconductors (0.17-0.3 eV) \cite{Hasan23, Shen23}.

To show the 2D band dispersion of the SS, we plot in Figs. 4(a)-(d) the ARPES intensity as a function of $k_x$ and $k_y$ for several binding energies.  As seen in (a), a ring-like intensity pattern with a small six-fold-symmetric modulation reflecting the symmetry of surface is clearly observed.  Upon approaching the Dirac point, the ring-like image gradually shrinks and converges into a single bright spot at the Dirac point (c), and then expands again below the Dirac point (d), like in graphene \cite{GrapheneDirac}.  Based on the observed energy position of the SS, we plot in Fig. 4(e) the experimental 2D band dispersion.  It is evident that the observed band dispersion forms a Dirac cone with an asymmetric shape in the energy axis with respect to the Dirac point.  The estimated band velocity at the Dirac point is 2.0 eV\AA (3.1$\times$10$^5$ m/s) along the $\bar{\Gamma}$$\bar{\rm{K}}$ direction, about 25$\%$ smaller than the value obtained for Bi$_2$Te$_3$ (2.7 eV\AA; 4.1$\times$10$^5$ m/s) \cite{Shen23}.

Now we discuss the nature of the observed SS.  In both bulk and surface, the TRS holds, requiring that the spin-dependent energy dispersion satisfies {\it E}({\it k}, $\uparrow$) = {\it E}(-{\it k}, $\downarrow$).  In addition, the space-inversion symmetry in the bulk dictates {\it E}({\it k}, $\uparrow$) = {\it E}(-{\it k}, $\uparrow$).Combining these two leads to {\it E}({\it k}, $\uparrow$) = {\it E}({\it k}, $\downarrow$), meaning that the spin degeneracy is not lifted in bulk.  On the other hand, on the surface where the space-inversion symmetry is broken, the TRS alone dictates the character of the bands.Therefore, if two bands are spin-split on the surface, they must show the Kramers degeneracy
\begin{figure}[h]
\includegraphics[width=3.3in]{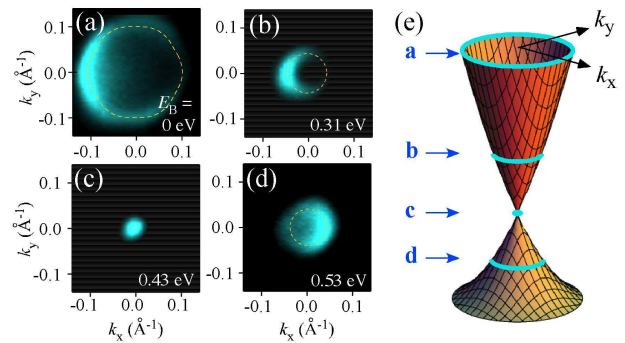}
\caption{(Color) (a)-(d) ARPES-intensity plot measured with Xe I line for different binding energies $E_{\rm B}$.   (e) 2D SS dispersion around the $\bar{\Gamma}$ point determined from the present ARPES data, together with the energy location of the 2D intensity maps (arrows a-d).}
\end{figure}
 at the time-reversal-invariant momenta (TRIMs) like the $\bar{\Gamma}$ point ($k_x$=$k_y$=0).  In the case of TlBiSe$_2$, the TRIMs are ƒ¡, L, F, and T points in the bulk BZ [see Fig. 1(b)], corresponding to the $\bar{\Gamma}$ and $\bar{\rm{M}}$ points in the surface BZ.  As clearly seen in Fig. 3, the present result definitely shows the degeneracy of the SS exactly at the $\bar{\Gamma}$ point, indicating that these SS are protected by the TRS \cite{KanePRL}, as observed in other surface Rashba systems \cite{SbSugawara, Au, Bi}.  We have extensively surveyed the band structure near $E_{\rm F}$ over the entire BZ and found no clear evidence for other SS near $E_{\rm F}$.   Therefore we conclude that the SS of TlBiSe$_2$ is characterized by a single Dirac cone at the BZ center.  This fact indicates an odd number of crossings between two TRIMs $\bar{\Gamma}$ and $\bar{\rm{M}}$ and an even number of crossings between independent $\bar{\rm{M}}$ points, satisfying the criteria for the strong TI with the topological index of $\nu$$_0$=1.  Hence, the present ARPES result gives evidence that TlBiSe$_2$ is a strong TI with the largest bulk energy gap to date.

Finally we comment on the relationship between the present ARPES result and the theoretical calculations in terms of the surface termination.   By repeating the sample cleavage several times, we have confirmed that the emergence of a single Dirac cone SS is robust to the cleavage condition, whereas it was theoretically predicted that a clean observation of a single Dirac cone would be possible only when the termination was between Se and Tl with the Se layer on the exposed surface, among the four possible cleavage planes \cite{Tl112Banzil}.  This may indicate that the cleaved surface in the ARPES measurements (typically 1$\times$1mm$^2$) has various (likely all possible) terminations owing to the absence of a natural cleavage plane.  It is also remarked that we find no indication of other SS away from the $\bar{\Gamma}$ point, unlike the calculation which predicts non-topological SS arising from dangling bonds \cite{Tl112Banzil}.  The origin of this discrepancy between experiment and theory is unclear at present and needs further investigations, possibly by employing real-space techniques such as the scanning tunneling microscopy.

In summary, we have reported the first ARPES study on ternary chalcogenide TlBiSe$_2$. We found compelling evidence for the Dirac-cone topological SS within the indirect bulk band gap.  The present result unambiguously indicates that TlBiSe$_2$ is a strong TI with a single Dirac cone at the BZ center.  The observed large bulk band gap of 0.35 eV provides a high potential for room-temperature spintronics applications.  A next important step would be to determine the spin structure of the Dirac cone as well as to achieve the truly bulk-insulating phase with $E_{\rm F}$ tuned to the Dirac point.

\begin{acknowledgments}
We thank K. Nakayama and T. Kawahara for their help in the ARPES measurement.  This work was supported by JSPS (KAKENHI 19674002), JST-CREST, MEXT of Japan,AFOSR (AOARD 10-4103), and KEK-PF (Proposal No. 2009S2-005).
\end{acknowledgments}

\end{document}